\documentclass[conference]{IEEEtran}
\usepackage{graphicx} 
\IEEEoverridecommandlockouts
\usepackage{cite}
\usepackage{amsmath,amssymb,amsfonts}
\usepackage{algorithmic}
\usepackage{graphicx}
\usepackage[font=footnotesize]{caption}
\usepackage{subcaption}
\usepackage{textcomp}
\usepackage{xcolor}
\usepackage{comment}
\usepackage{balance}
\usepackage{float}
\usepackage[export]{adjustbox}
\def\BibTeX{{\rm B\kern-.05em{\sc i\kern-.025em b}\kern-.08em
    T\kern-.1667em\lower.7ex\hbox{E}\kern-.125emX}}

\begin{document}
\title{Distributed Intelligent Integrated Sensing and Communications: The 6G-DISAC Approach
\thanks{This work has been supported by the SNS JU project 6G-DISAC under the EU's Horizon Europe research and innovation program under Grant Agreement No 101139130.}
}

\author{Emilio Calvanese Strinati\IEEEauthorrefmark{1}, George C. Alexandropoulos\IEEEauthorrefmark{2}, Madhusudan Giyyarpuram\IEEEauthorrefmark{3},\\ Philippe Sehier\IEEEauthorrefmark{4}, Sami Mekki\IEEEauthorrefmark{4}, Vincenzo Sciancalepore\IEEEauthorrefmark{5}, Maximilian Stark\IEEEauthorrefmark{6}, Mohamed Sana\IEEEauthorrefmark{1},\\ Benoit Denis\IEEEauthorrefmark{1}, Maurizio Crozzoli\IEEEauthorrefmark{7}, Navid Amani\IEEEauthorrefmark{8}, Placido Mursia\IEEEauthorrefmark{5}, Raffaele D'Errico\IEEEauthorrefmark{1},\\ Mauro Boldi\IEEEauthorrefmark{7}, Francesca Costanzo\IEEEauthorrefmark{1}, Francois Rivet\IEEEauthorrefmark{9}, and Henk  Wymeersch\IEEEauthorrefmark{10}
\\
\IEEEauthorrefmark{1}CEA Leti, France
\IEEEauthorrefmark{2}National and Kapodistrian University of Athens, Greece\\
\IEEEauthorrefmark{3}Orange Innovation, France
\IEEEauthorrefmark{4}Nokia Bell Labs, France
\IEEEauthorrefmark{5}NEC Laboratories Europe, Germany\\
\IEEEauthorrefmark{6}Robert Bosch, Germany
\IEEEauthorrefmark{7}TIM Telecom Italia, Italy
\IEEEauthorrefmark{8}RadChat, Sweden\\
\IEEEauthorrefmark{9}Univ. Bordeaux, France
\IEEEauthorrefmark{10}Chalmers University of Technology, Sweden
}

\maketitle

\begin{abstract}
This paper introduces the concept of Distributed Intelligent integrated Sensing and Communications (DISAC), which expands the capabilities of Integrated Sensing and Communications (ISAC) towards distributed architectures. 
Additionally, the DISAC framework integrates novel waveform design with new semantic and goal-oriented communication paradigms, enabling ISAC technologies to transition from traditional data fusion to the semantic composition of diverse sensed and shared information. This progress facilitates large-scale, energy-efficient support for high-precision spatial-temporal processing, optimizing ISAC resource utilization, and enabling effective multi-modal sensing performance.
Addressing key challenges such as efficient data management and connect-compute resource utilization, 6G-DISAC stands to revolutionize applications in diverse sectors including transportation, healthcare, and industrial automation. Our study encapsulates the project’s vision, methodologies, and potential impact, marking a significant stride towards a more connected and intelligent world.
\end{abstract}


\section{Introduction}
The swift advancement in wireless communication technologies has ushered networks from the 4G era to 5G, enhancing mobile broadband connectivity, reducing latency, and scaling device connections. The emergence of 6G represents a revolutionary shift, expanding the function of radio signals beyond communications to include radio sensing and environmental mapping capabilities. In this evolving landscape, the concept of Integrated Sensing and Communications (ISAC) plays a critical role in the development of the 6G wireless networks~\cite{LHL22,DA23}. ISAC in 6G is anticipated to mark the beginning of a new connectivity era, where communication transcends mere data transfer and incorporates sensing, intelligence, and reconfigurability\cite{ETSIISG,3GPPTR22837}. This integration bridges the physical and digital realms~\cite{DT_RIS}, redefining our interaction and experience of the surrounding world.

The current vision for ISAC often overlooks several essential elements, e.g., support for widespread deployments to monitor numerous connected User Equipments (UEs) and passive objects over vast areas and time periods. In addition, it is envisioned to integrate 6G signals with external sensors and broader semantic awareness~\cite{CalvaneseGOWSC2021}, rather than solely relying on the 6G signal as a sensor.  To harness the full potential of ISAC, it is crucial to expand its vision by addressing these  critical aspects. This paper introduces such an expanded vision via the 6G-DISAC project. The 6G-DISAC project envisions a transformative approach to wireless connectivity, enhancing both internal 6G operations and external services and applications. It amalgamates diverse sensor fusion with an adaptive, efficient, semantic-native strategy, enabling energy-efficient and high-resolution tracking of both connected UEs and objects. By integrating sensing with communications and leveraging distributed Artificial Intelligence (AI)~\cite{DAI_6G}, the 6G-DISAC project, not only intends to overcome existing ISAC limitations, but also aims to unlock new possibilities for efficient, precise, and semantic operations.

The foundation of 6G-DISAC is to build on interrelated key components. The first component is a \textit{semantic and goal-oriented} framework supported by Machine Learning (ML) and AI. This intelligent and efficient framework encompasses various elements, such as sensing activation, waveform design, physical (PHY) layer components, resource allocation, robust protocols, and multi-modal semantic reasoning. It ensures outstanding sensing performance across numerous applications while optimizing resource usage. The second component, the \textit{6G-DISAC architecture}, underpins both sensing and communications, facilitating intelligent operations and distributed functions~\cite{distributed_TX_interference,distributed_RIS_sum,distributed_RIS_loc}. This architecture revolutionizes the fabric of wireless networks by enabling extensive tracking and supporting distributed AI operations~\cite{Alexandropoulos2022Pervasive}. It balances local data processing and fusion, incorporates novel multi-antenna technologies~\cite{FD_MIMO,RISoverview2023}, and integrates an exposure framework for external sensors. The third component focuses on \textit{advanced high-resolution processing}, leveraging massively distributed observations and balancing computational and storage demands. Efficient methods combining ML and model-based signal processing must be developed and validated to support this framework over the 6G-DISAC architecture.

In the following sections, this paper delves into the 6G-DISAC project, covering the key DISAC architectural components, the main technology enablers, and proof of concepts.

\section{Distributed ISAC System Architecture}
In this section, we discuss DISAC use cases, their quantitative targets, and key architectural components.
\subsection{Scenario and Use Cases}
The potential applications of the 6G-DISAC project are diverse and impactful, particularly when viewed through the lens of its key components: the semantic framework, the architecture, and high-resolution processing. These components not only enhance the capabilities of 6G-DISAC, but also align closely with the needs of various use cases.

\begin{figure}[t]
    \centering
    \includegraphics[width=0.8\columnwidth]{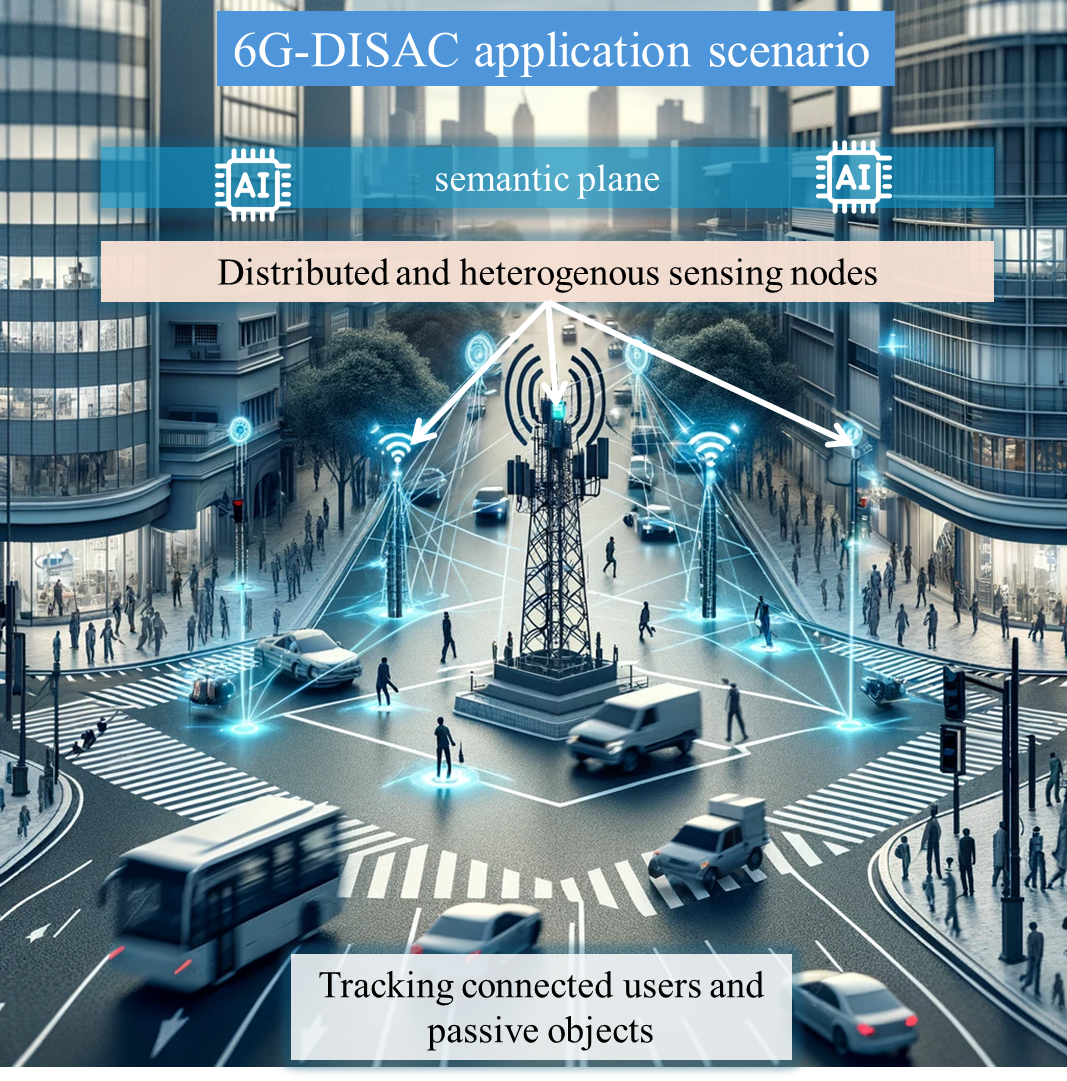}
    \caption{An example 6G-DISAC application scenario (urban digital twin), providing the ability to track users and objects over extended time and space, supported by the DISAC architecture and the integrated semantic framework.}
    \vspace{-4mm}
    \label{fig:DISAC-scenario}
\end{figure}
\textbf{Intelligent Transportation:} The 6G-DISAC project can significantly enhance traffic monitoring and the protection of vulnerable road users, such as pedestrians and cyclists. The semantic framework of 6G-DISAC, with its goal-oriented and intelligent design, can dynamically adapt to different traffic scenarios, optimizing the allocation of sensing and communication resources for real-time traffic analysis. The architecture of 6G-DISAC, with its distributed nature, will enable extensive coverage, ensuring that even passive targets, like unconnected vehicles or pedestrians, are detected and tracked over an area of interest. High-resolution processing capabilities will allow for precise measurements of position, speed, and direction, enhancing the accuracy of detection and prediction of traffic flows and potential hazards. In combination with an Urban Digital Twin (see Fig.~\ref{fig:DISAC-scenario}), this provide an invaluable tool for urban planners and policymakers, providing real-time insights into the dynamics of the cityscape for informed decision-making and sustainable urban development. 


\textbf{Smart Factory/City:} Industrial applications, including presence and intrusion detection, geofencing, and gesture recognition for digital twins, will benefit greatly from the 6G-DISAC project. The envisioned semantic framework will enable the system to understand and interpret complex industrial environments, distinguishing between routine operations and anomalies and the composition of semantically selected information, as well as the pragmatic generation of AI reasoning stimuli \cite{ThomasCCNC2024}. This capability is crucial for maintaining security and operational efficiency. In terms of anti-collision systems for autonomous ground vehicles and unmanned aerial vehicles detection, the high-resolution processing component of 6G-DISAC plays a pivotal role. It will allow for precise tracking and prediction of vehicle movements, reducing the risk of collisions and enhancing safety. Furthermore, air quality analysis, in sectors like agriculture, will benefit from the distributed architecture, which allows for the collection and analysis of environmental data over large areas, aiding in more accurate and timely decision-making for crop management.

In addition to the highlighted applications, the 6G-DISAC project enables various other potential use cases outlined by industrial fora like NGMN, 5GAA, and 5G-ACIA.

\begin{figure*}[t]
    \begin{center}
\includegraphics[width=0.88\textwidth, center]{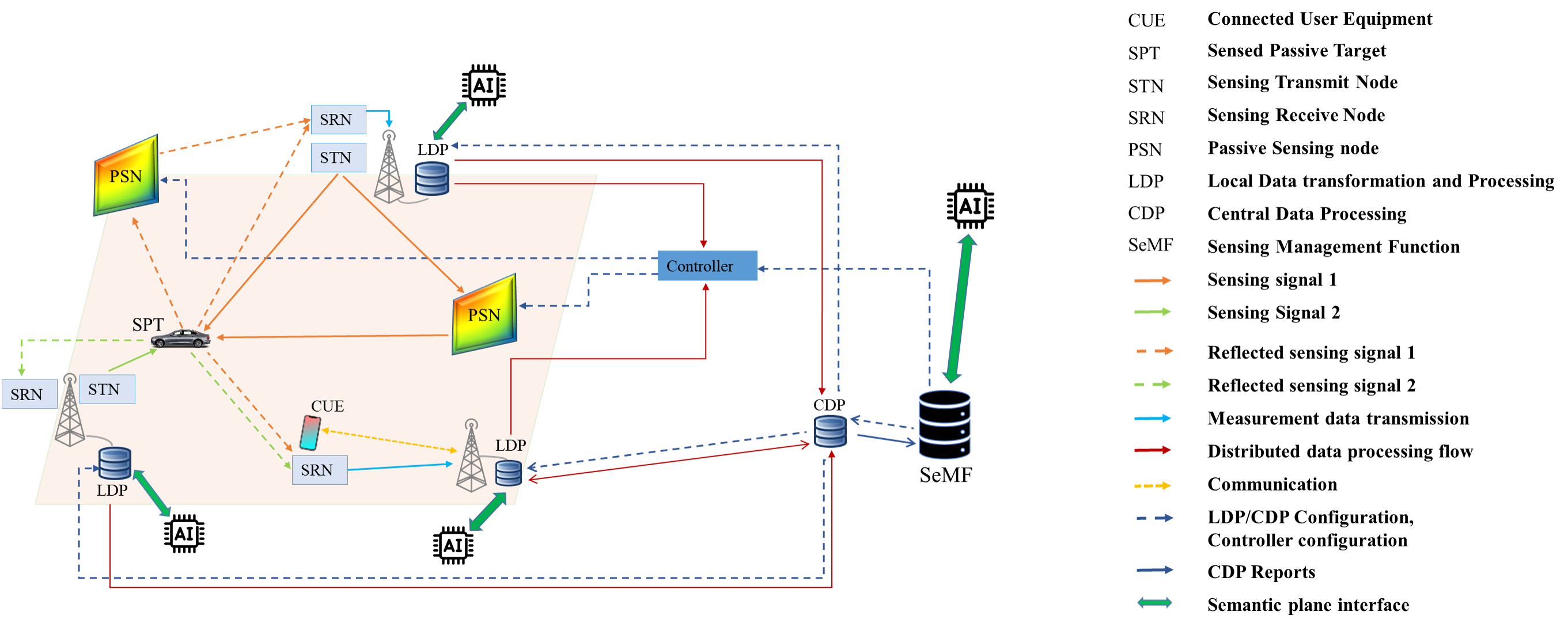}
    \end{center}
\caption{The vision for the 6G-DISAC network architecture including the SeMF component tasked to configure and manage STNs and SRNs.} 
\label{fig:arch}
\end{figure*}
\subsection{Metrics and Key Performance Indicators (KPIs)}
The DISAC framework will optimize ISAC PHY-layer and system performance, focusing on waveform shaping, hardware compliance, and new signal and semantic processing techniques. Metrics and KPIs for distributed ISAC and resource allocation trade-offs between communication and sensing will be crucial for evaluating if specific ISAC use case KPIs are achieved.
This requires to go beyond the usual metrics defined in the single sensing receive node case to consider scenarios where multiple nodes are involved in a collaborative manner to perform sensing. The DISAC framework can be evaluated by seven distinct Quantitative Targets (QTs), each addressing a specific aspect of technological advancement.

\noindent 
{\bf QT1: Spectral Efficiency.} Aiming for a relative improvement of $30\%$ over the current 3GPP R18 Multiple-Input Multiple-Output (MIMO) benchmarks. This enhancement is anticipated to be enabled through sensing-aided communications (e.g.,~\cite{FD_MIMO,RISoverview2023}), which is expected to refine the estimation of MIMO channel matrices and geometrical channel parameters, thereby elevating spectral efficiency.

{\noindent \bf QT2: User Positioning Accuracy.} Unlike the current undefined baseline, the target is set at an absolute value of $0.01$ m in Frequency Range 2 (FR2) and $0.1$ m in Frequency Range 1 (FR1). This precise positioning is achievable through the combination of positioning and sensing, alongside the utilization of large bandwidths and apertures.

{\noindent \bf QT3: Energy Efficiency.} DISAC aims for a $40\%$ improvement in FR2, compared to the standards outlined in 3GPP TR 38.684. Similar to QT1, this goal is expected to be reached through sensing-aided communication, which will enhance the estimation of geometrical channel parameters, thus contributing to greater energy efficiency.

{\noindent \bf QT4: User Orientation Accuracy.} DISAC seeks to achieve an accuracy of less than $1^{\circ}$ in FR2, an improvement from the current baseline of $1^{\circ}-5^{\circ}$. This heightened accuracy is anticipated through the use of antenna array technologies at the UE side and the involvement of numerous Base Stations (BSs) / access points.

{\noindent \bf QT5: Object Positioning Accuracy.} DISAC targets for less than $1$ m in FR1 and $0.1$ m in FR2, improving upon the current baseline of $1$ m. This is envisioned to be achieved through the use of large apertures in FR1 and large bandwidths and arrays in FR2, in combination with numerous transmitters and receivers in both frequency ranges.

{\noindent \bf QT6: Object Velocity Estimation Accuracy.} The targeted  accuracy is less than $1$ m/s in FR1, an improvement over the current baseline of $1$ m/s. This target is based on the potential for large integration times in FR1.

\subsection{Architectural Proposal}
The 3GPP architecture today offers support for varying communication requirements (eMBB, URLLC, and mMTC) and the positioning of UEs. However, significant changes are needed to fully integrate sensing in the architecture. 6G-DISAC's goal is to design a network architecture that can detect multiple targets using multiple sensing nodes, spread over a geographical area of interest, for different applications. This architecture needs to be integrated with 3GPP compliant systems, that are communications-wise optimized, to provide seamless communications and sensing using the same infrastructure and resources. 
Figure \ref{fig:arch} illustrates a high-level overview of the envisioned 6G-DISAC network architecture, where several new entities are introduced: the \textit{sensing targets}, i.e., the object(s) that need to be sensed. The designed architecture will address distributed sensing and communications involving multiple targets that can be passive or active. Those targets will be spread over several cells, and will involve multiple Sensing Transmit Nodes (STNs), Sensing Receive Nodes (SRNS), and Passive Sensing Nodes (PSNs). The distributed intelligent signal processing will enable the collaboration between these sensing nodes to provide a multi-perspective and integrated view of the environment of interest. This entails, not just fusion of data from multiple SRNs, but also a map of the environment. Given the potentially huge amount of data generated by the distributed sensing, data will necessitate local processing at the SRNs, so that only higher-level sensing information is sent to the fusion center. Several Local Data Processing (LDPs) entities will collaborate to perform data transformation to ensure that the communication capacity of the system is not impacted. However, due to the heterogeneous nature of the devices participating in the DISAC architecture, the level of compression achievable at different nodes will vary according to device capabilities. Hence, a flexible  data model is needed for the exchange of information with the fusion center.  
6G-DISAC will work over an extended geographical area and over time. Thus, passive and active objects need to be tracked over space and time. New functions will also need to be introduced in the architecture, dedicated to tracking and handover of objects. This, in turn, requires the definition and management of SRN groups, as well as suitable identifiers for objects (an object will not possess a Subscriber Identity Module (SIM) card). In addition, 6G-DISAC will provide semantic communications between different network functions, which adds new network functions and protocols. Those functions include semantic extraction of information using the context of communications, semantic composition of information from various sources, and semantic instruction to the sensing device. Among the distributed functions that need to be supported by the network architecture, the relevant AI modules at the semantic layer (but also those at the radio layer) stand out in this regard. For training, inference, and exchange of information to enable federated and multi-agent learning, functions, and interfaces must be provided. The architecture will also make use of different types of Reconfigurable Intelligent Surfaces (RISs)~\cite{RISoverview2023} to enhance both communications and sensing. This includes passive RISs, transparent to the architecture, that need to be managed, and RISs endowed sensing capabilities, that can act as an SRN with limited local processing power. 

\section{Research on Innovative Technology Enablers} 
\subsection{Waveforms and Semantic Communications}
\begin{figure}[!t]
    \centering
    \includegraphics[width=0.95\columnwidth]{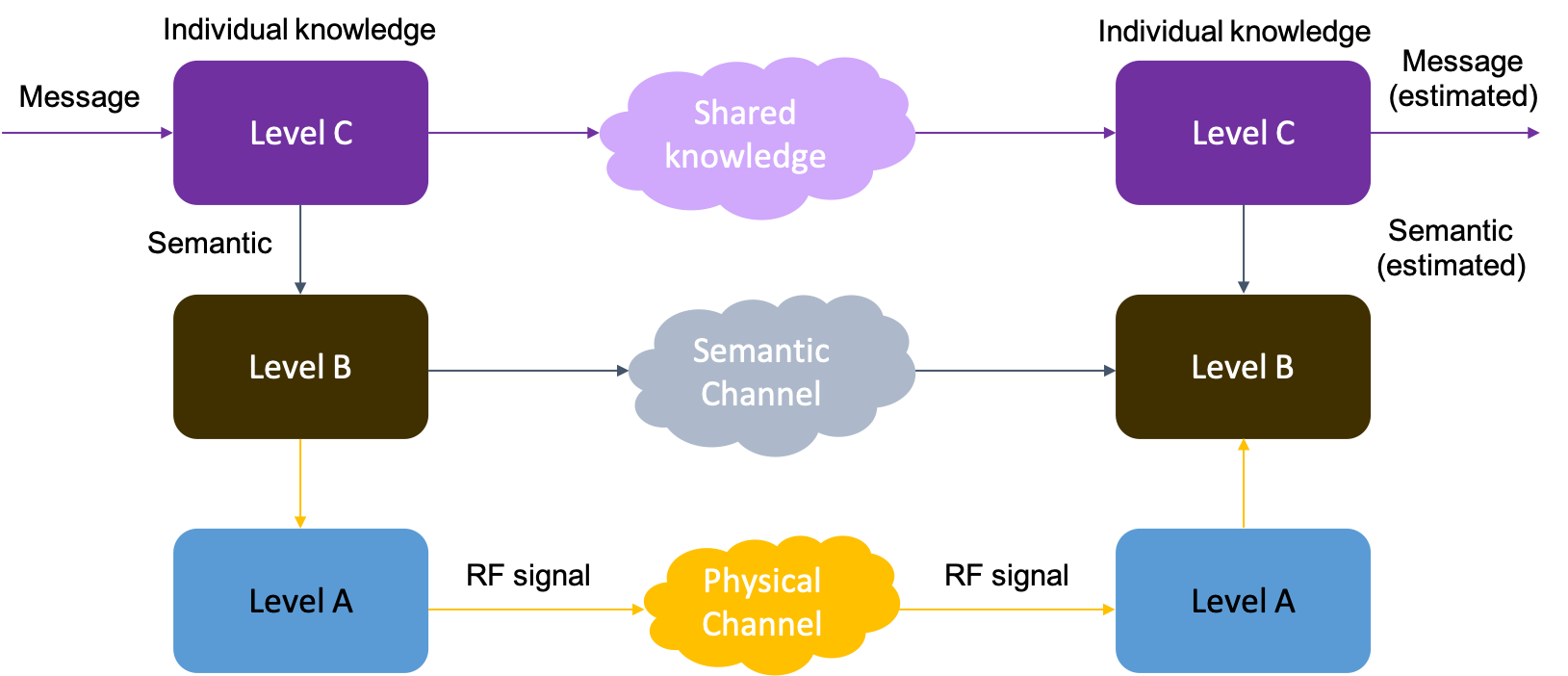}
    \caption{Levels of information sharing in 6G-DISAC.} \vspace{-2mm}
    \label{fig:semantic}
\end{figure}

The 6G-DISAC project is set to optimize the performance of DISAC systems through a novel, comprehensive design approach. This approach will incorporate new signal processing techniques, focusing on innovative waveform shaping and supporting hardware components. Key areas of development include designing waveforms that excel in both sensing and communications and operate efficiently in both FR1 and FR2 frequency ranges. Those waveforms will facilitate semantic communications, utilizing information about the environment and mutual information exchange among networks nodes for enhanced data conveyance. Conventional communications focuses on the transmission of bits over the physical radio channel, while semantic communications in the 6G-DISAC project will also consider the semantic channel as well as the higher level sharing of knowledge (see Fig.~\ref{fig:semantic}). 

The methodology of 6G-DISAC will optimize hardware for power efficiency and waveforms for spectral efficiency, incorporating AI/ML-based semantic-aware solutions. This aims to improve task-level performance in localization, sensing, communications, and control, while utilizing fewer resources. The project aims to surpass current technological limits by adapting waveforms for both wireless communications and semantic channels, enhancing ISAC environments.


A key innovation in 6G-DISAC is the design of semantic-aware waveforms \cite{patent_SemComsECS-FR}. These waveforms will be structured to perform both wideband sensing and specific encoding tasks simultaneously. The design process considers various waveform types (temporal, sequential, and frequential) and utilizes coefficients for specific semantic encoding (see definitions in \cite{CalvaneseGOWSC2021}). These coefficients are divided into three clusters, each with a specific purpose: sensor pilots for sensing, semantics for encoding information, and resources for error correction to improve communication robustness and sensing accuracy.
This innovative waveform shaping in 6G-DISAC will consider several trade-offs to optimize hardware use, focusing on energy costs, receiver robustness, sensing accuracy, and spectral occupation. The project aims to leverage semantic communication principles to relax constraints at the radio transmission level, allowing for more efficient and robust wireless communications and sensing. The expected outcomes include improved sensor data management, enhanced deployment flexibility of sensing infrastructure, and robustness against wireless channel impairments.

\subsection{Intelligent and Distributed PHY Layer}
Designing the PHY layer for the 6G-DISAC concept presents significant challenges, requiring strategic resource allocation to support both communications and sensing functions. This entails optimizing radio resources and positioning BSs, RISs, Distributed-MIMO (DMIMO), and Extremely-Large MIMO (XL-MIMO) access points. Resource allocation differs between communication and sensing, with the latter relying on multi-point connections. The project also faces the challenge of efficiently processing large volumes of data, particularly in multi-sensor and multi-target scenarios. This includes developing novel signal processing techniques to align with optimized waveforms and addressing complex data association issues. Additionally, local processing, compression, and information sharing are vital in the context of the semantic-aware DISAC system.

Another critical challenge is to enhance communication performance. Although sensing and positioning consume radio resources, they provide a detailed view of the environment, connected UEs, passive objects, and their trajectories. This rich information, particularly in a distributed setting, can be leveraged to improve communications-oriented KPIs, especially with large MIMO architectures and low-power sensing devices. To address this challenge, 6G-DISAC aims to develop scalable, adaptive, and distributed resource allocation methods, considering various deployment options (conventional BSs, D-MIMO, XL-MIMO, passive RISs, and multi-functional RISs). New signal processing methods for target detection, estimation, and tracking will focus on distributed measurements and inter-device cooperation. Large MIMO architectures, including RIS and hybrid RISs~\cite{alexandropoulos2023hybrid}, will be incorporated into all PHY-layer aspects. ML and AI will play a crucial role in complex optimization and design problems, inspiring model-based solutions. Federated learning and distributed auto-encoders will be also utilized for learning waveforms for distributed transmissions, receiver-side signal processing, optimized deployments, as well as resource allocation.

Semantic communications will also be a focus, with a semantic channel learned between transmitters and receivers to facilitate smart sensor activation, minimal resource consumption during sensing, and minimal information sharing for task-oriented KPIs. When needed, if the semantic noise resulting from critical distortion introduced by semantic channels, compensation techniques such as semantic channel equalization can be applied \cite{sana2023}. The use of RISs will be explored from both model-driven and ML perspectives~\cite{Alexandropoulos2022Pervasive}, examining implications on waveform design, resource allocation, and signal processing. Programmable reflective capabilities of RISs will be exploited for ISAC, enhancing the correlation between communication and sensing channels.

Finally, communications will serve both as an enabler and a service of DISAC. Large-scale and distributed communication provisioning will be optimized with location and context information, encompassing full-duplex MIMO systems~\cite{FD_MIMO}, D-MIMO, massive MIMO, and RISs. The availability of sensing information will be used to optimize analog and digital precoders and combiners in MIMO systems, targeting various KPIs, such as interference handling, privacy and security, and unintended ElectroMagnetic Field (EMF) exposure. Advanced localization techniques and context-aware solutions will be integrated to improve communication efficiency, leading to enhanced user experiences and more reliable connections.

\begin{figure*}
    \centering
    \includegraphics[width=0.85\textwidth]{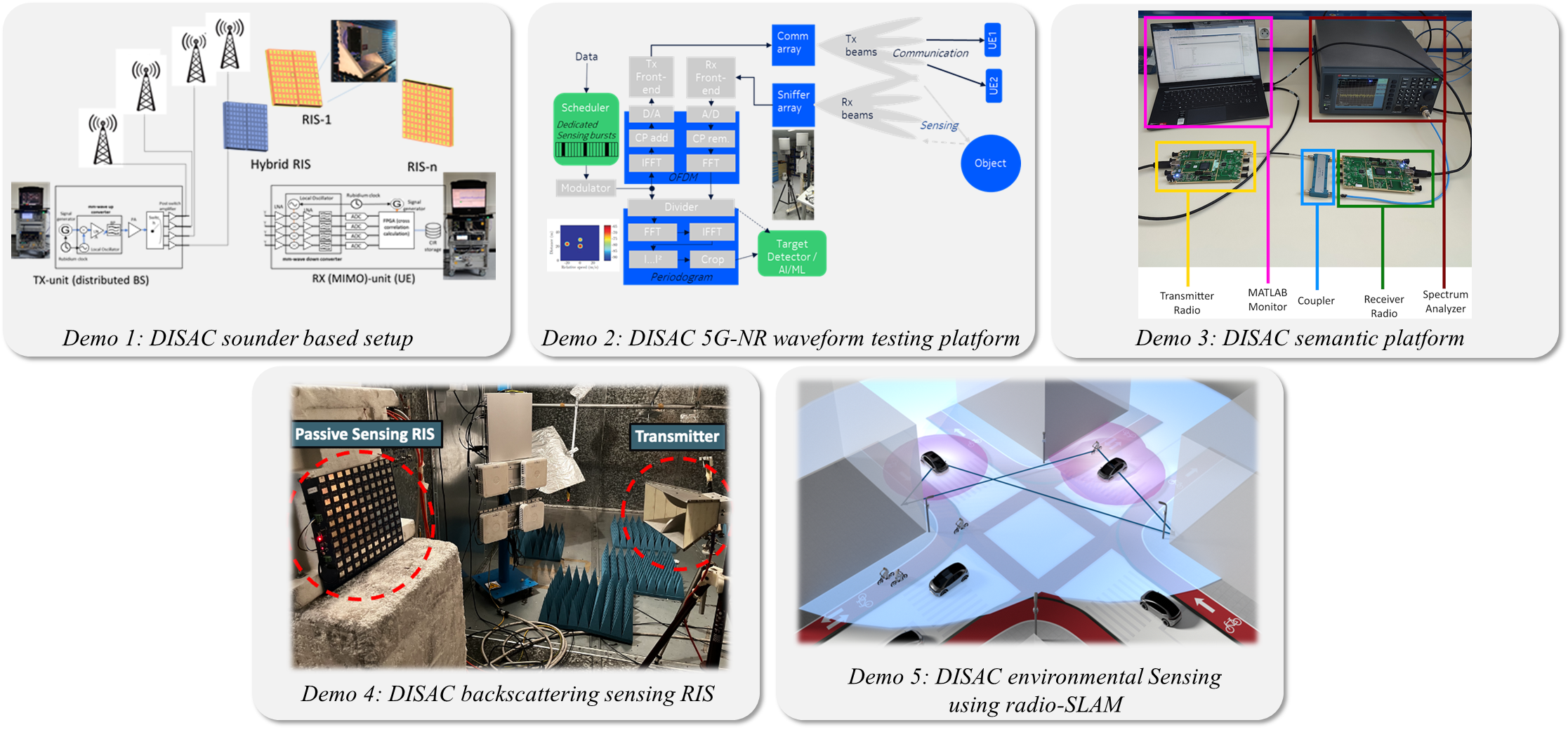}
    \caption{Overview of the five planned 6G-DISAC demonstrations.}\vspace{-5mm}
    \label{fig:DISAC-demo}
\end{figure*}

\section{DISAC Proofs of Concept} 
The 6G-DISAC project, through its innovative distributed architecture, aims to validate and refine the concepts to be developed by comparing and selecting the best methods from a variety of demonstrators.

\subsection{Description of the 6G-DISAC Demos}
Five demonstrations are planned, covering four testing platforms and one comprehensive trial, as illustrated in Fig.~\ref{fig:DISAC-demo}. They will collectively showcase the advanced capabilities of the envisioned DISAC system. 
 %

\emph{\textbf{Demo 1: DISAC sounder based setup.}}
This demo will include a multi-purpose setup for multi-point measurements, utilizing a channel sounder and an RIS at FR1 and FR2 frequency bands. XL-MIMO will be realized through a virtual array and DMIMO (with up to four different BS locations) will be implemented through optical fiber / Radio-Frequency (RF) extension from the central unit. Various UE mobility regimes will be tested through in-lab experiments and indoor/outdoor scenarios. The data collected in FR1 and FR2 bands will be processed using high-resolution algorithms developed during the project for distributed communications and sensing, along with AI-based methods, like distributed auto-encoders and graph signal processing, enabling advanced UE tracking and sensing, with emphasis on semantic and goal-oriented sensing.

\emph{\textbf{Demo 2: DISAC 5G-NR waveform testing platform.}}
This demo will focus on collecting real-life measurements using Nokia’s equipment. In particular, Nokia's 5G gNB product operating at $27$ GHz in various configurations will be deployed, including monostatic and multi-static setups. The 5G NR waveform will be used for sensing and communications. The configuration flexibility of this waveform, supported by the gNB, will be exploited to test several modes combining sensing and communications. Multi-static sensing will be emulated by performing measurements with different transmitter and receiver positions. Tests involving RISs will be also considered with compatible RIS prototypes. Real-time as well as offline processing will be implemented. This demo will concentrate on efficient resource utilization for developing distributed sensing alongside communications. 

\emph{\textbf{Demo 3: DISAC semantic platform.}}
This demo features a radio set utilizing semantic-aware waveforms for communications and sensing, aiming to validate adaptive waveform shaping for ISAC across various channel topologies. Initial assessments will be conducted using software-defined radios (SDRs) at $2$ GHz, followed by application-specific RF integrated circuits designed by CEA Leti and University of Bordeaux at $5$ GHz, with access to FR2. A benchmark will evaluate ISAC performance across different channel topologies and confirm waveform shaping selection based on semantic communications. Both wireless and semantic channel metrics \cite{Getu23SEmMetrics} will be used to benchmark the added value brought using semantic-aware waveforms designs.

\emph{\textbf{Demo 4: DISAC backscattering sensing RIS.}}
This demo will explore the integration of backscattering and RISs for joint communications and sensing. Operating in the ${\rm S}$ and ${\rm Ka}$ bands, this setup will incorporate a low-power, almost passive RIS, and AI/ML-based orchestration solutions, to incorporate sensing and AI/ML as main ingredients of a 6G network's intelligence. Both passive and active sensing nodes will be deployed using the RIS technology as lightweight and low complexity devices. This demonstrator will be comprised: \textit{i}) backscatter ISAC by exploiting active RISs, i.e., RISs that are fed with a power source; and \textit{ii}) RISs with nearly passive electronic components, which require only a negligible power input, and, moreover, integrate limited sensing capabilities in the form of a directional power detector. The goal of this task is to demonstrate, not only the feasibility of joint communications and sensing at the link-layer level, but also to show that the system performance can be improved by an adequate use of sensing RISs. Adaptation and optimization of the transmission parameters to different scenarios and configurations will also be shown by means of this demonstration campaign. 

\emph{\textbf{Demo 5: DISAC environmental sensing using radio-SLAM.}}
This demo will build on the outcomes of Demos $1-4$ to focus on sensing multiple targets in a distributed configuration. Information exchanged between multiple UEs will be fused to enhance position accuracy and perform environmental sensing, leading to the creation of environmental maps and the realization of distributed radio-SLAM. The idea of intelligent distributed sensing and communication will be addressed on high layers. Objects will be sensed, and the information will be exchanged over the sidelink in FR1. 
In addition, UEs will be considered as active users, i.e., they can communicate. As a result, dedicated positioning reference signals (e.g., Positioning Reference Signals (PRSs) in 5G NR R17) will be transmitted between the UEs that will be used for positioning. Nevertheless, while performing positioning, also the environment can be sensed. As a result, a map of the environment will be constructed, while performing the positioning task. Especially, in rich multipath environments, distributed radio-SLAM and cooperation can be beneficial to improve the localization accuracy of all involved participants. 


\subsection{Discussion}
%
Regarding the Technology Readiness Level (TRL), the Proof of Concepts (PoCs) in the latter panned demonstrations will support multi-static configurations with distributed sensing capabilities in indoor and outdoor settings. By the project's end, the benefits of distributed sensing in enhancing target detection and identification performance, including advancements in localization accuracy, speed, and target tracking, will be showcased. The integration of sensing and communications will be demonstrated under various conditions, including with and without RIS and semantics, highlighting the reciprocal benefits of sensing information to communications, and vice versa. The TRL at this stage is expected to range between $4$ and $5$, indicating significant progress from theoretical research to practical applications.

\section{Conclusion}
The envisioned DISAC framework extends the ISAC concept to distributed and intelligent implementations, integrating advanced waveform design and semantic communications with advanced PHY-layer components.The proposed framework will build on three main pillars: (1) an AI-native distributed architecture for ISAC; (2) semantic and goal-oriented communication paradigms, and (3) high-resolution sensing. This novel approach enables the transition from classical data fusion to the composition of semantically selected information. Critical to its success are the challenges of optimally allocating resources for concurrent communications and sensing, efficiently processing vast amounts of data and semantically composing rich environmental information to improve communications performance. Additionally, specific metrics and KPIs tailored to DISAC will be developed, along with appropriate channel models, to assess performance in relevant use cases.
The DISAC framework will develop scalable and adaptive solutions integrating cutting-edge AI/ML techniques, thus, paving the way for revolutionary applications in various sectors.

\balance 
\bibliographystyle{IEEEtran}
\bibliography{bibliography}

\begin{thebibliography}{10}
\providecommand{\url}[1]{#1}
\csname url@samestyle\endcsname
\providecommand{\newblock}{\relax}
\providecommand{\bibinfo}[2]{#2}
\providecommand{\BIBentrySTDinterwordspacing}{\spaceskip=0pt\relax}
\providecommand{\BIBentryALTinterwordstretchfactor}{4}
\providecommand{\BIBentryALTinterwordspacing}{\spaceskip=\fontdimen2\font plus
\BIBentryALTinterwordstretchfactor\fontdimen3\font minus \fontdimen4\font\relax}
\providecommand{\BIBforeignlanguage}[2]{{%
\expandafter\ifx\csname l@#1\endcsname\relax
\typeout{** WARNING: IEEEtran.bst: No hyphenation pattern has been}%
\typeout{** loaded for the language `#1'. Using the pattern for}%
\typeout{** the default language instead.}%
\else
\language=\csname l@#1\endcsname
\fi
#2}}
\providecommand{\BIBdecl}{\relax}
\BIBdecl

\bibitem{LHL22}
A.~Liu \emph{et~al.}, ``A survey on fundamental limits of integrated sensing and communication,'' \emph{IEEE Commun. Surveys \& Tuts.}, vol.~24, no.~2, pp. 994--1034, 2022.

\bibitem{DA23}
U.~Demirhan and A.~Alkhateeb, ``Integrated sensing and communication for {6G}: Ten key machine learning roles,'' \emph{IEEE Commun. Mag.}, vol.~61, no.~5, p. 113–119, May 2023.

\bibitem{ETSIISG}
ETSI, ``Integration sensing and communication industry specification group,'' ETSI, Tech. Rep., 2023.

\bibitem{3GPPTR22837}
3GPP, ``Study on integrated sensing and communication, release 19,'' 3GPP SA1, Tech. Rep., 2023.

\bibitem{DT_RIS}
A.~Masaracchia \emph{et~al.}, ``Towards the metaverse realization in {6G}: Orchestration of {RIS}-enabled smart wireless environments via digital twins,'' \emph{IEEE Internet of Things Mag.}, early access, 2023.

\bibitem{CalvaneseGOWSC2021}
E.~Calvanese~Strinati and S.~Barbarossa, ``{6G} networks: Beyond {S}hannon towards semantic and goal-oriented communications,'' \emph{J. Commun. Netw.}, vol. 190, Feb. 2021.

\bibitem{DAI_6G}
S.~Talwar \emph{et~al.}, ``{6G}: Connectivity in the era of distributed intelligence,'' \emph{IEEE Commun. Mag.}, vol.~59, no.~11, pp. 45--50, Nov. 2021.

\bibitem{distributed_TX_interference}
G.~C. Alexandropoulos and C.~B. Papadias, ``A reconfigurable distributed algorithm for {$K$}-user {MIMO} interference networks,'' in \emph{Proc. IEEE ICC}, Budapest, Hungary, 2013.

\bibitem{distributed_RIS_sum}
K.~D. Katsanos \emph{et~al.}, ``Distributed sum-rate maximization of cellular communications with multiple reconfigurable intelligent surfaces,'' in \emph{Proc. IEEE SPAWC}, Oulu, Finland, 2022.

\bibitem{distributed_RIS_loc}
J.~He \emph{et~al.}, ``Compressed-sensing-based {3D} localization with distributed passive reconfigurable intelligent surfaces,'' in \emph{Proc. IEEE ICASSP}, Rhodes, Greece, 2023.

\bibitem{Alexandropoulos2022Pervasive}
G.~C. Alexandropoulos \emph{et~al.}, ``Pervasive machine learning for smart radio environments enabled by reconfigurable intelligent surfaces,'' \emph{Proc. IEEE}, vol. 110, no.~9, pp. 1494--1525, Sep. 2022.

\bibitem{FD_MIMO}
B.~Smida \emph{et~al.}, ``Full-duplex wireless for {6G}: Progress brings new opportunities and challenges,'' \emph{IEEE J. Sel. Areas Commun.}, vol.~41, no.~9, p. 2729–2750, Sep. 2023.

\bibitem{RISoverview2023}
E.~Basar \emph{et~al.}, ``Reconfigurable intelligent surfaces for {6G}: {E}merging hardware architectures, applications, and open challenges,'' \emph{arXiv preprint arXiv:2312.16874}, 2023.

\bibitem{ThomasCCNC2024}
C.~T.~Kurisummoottil \emph{et~al.}, ``Reasoning with the theory of mind for pragmatic semantic communication,'' in \emph{Proc. IEEE CCNC}, Las Vegas, USA, Jan. 2024.

\bibitem{patent_SemComsECS-FR}
E.~Calvanese~Strinati \emph{et~al.}, ``System, method and computer program product for communicating semantic messages over a communication channel,'' Patent 23\,305\,542.5, April 12, 2023.

\bibitem{alexandropoulos2023hybrid}
G.~C. Alexandropoulos \emph{et~al.}, ``Hybrid reconfigurable intelligent metasurfaces: Enabling simultaneous tunable reflections and sensing for {6G} wireless communications,'' \emph{IEEE Veh. Technol. Mag.}, to appear, 2024.

\bibitem{sana2023}
M.~Sana and E.~Calvanese~Strinati, ``Semantic channel equalizer: Modelling language mismatch in multi-user semantic communications,'' in \emph{Proc. IEEE GLOBECOM}, Kuala Lumpur, Malaysia, 2023.

\bibitem{Getu23SEmMetrics}
T.~M. Getu \emph{et~al.}, ``Making sense of meaning: A survey on metrics for semantic and goal-oriented communication,'' \emph{IEEE Access}, vol.~11, pp. 45\,456--45\,492, 2023.

\end{thebibliography}

\end{document}